\documentstyle[aps,epsbox]{revtex}
\tighten

\begin{document}

\draft


\preprint{CITA-99-44, DAMTP-1999-177, RESCEU-50/99, UTAP-357, 
hep-th/9912287}
\title{Global structure of exact cosmological solutions in the brane 
world}
\author{Shinji Mukohyama}
\address{
Canadian Institute for Theoretical Astrophysics, 
University of Toronto\\
Toronto, ON, M5S 3H8\\
and \\
Department of Physics and Astronomy, 
University of Victoria\\ 
Victoria, BC, Canada V8W 3P6
}
\author{Tetsuya Shiromizu}
\address{
DAMTP, University of Cambridge \\ 
Silver Street, Cambridge CB3 9EW, United Kingdom\\
Department of Physics, The University of Tokyo, Tokyo 113-0033, Japan\\
and \\
Research Centre for the Early Universe(RESCEU), \\ 
The University of Tokyo, Tokyo 113-0033, Japan}
\author{Kei-ichi Maeda}
\address{
Department of Physics, Waseda University, Shinjuku, 
Tokyo 169-8555, Japan
}
\date{\today}

\maketitle


\begin{abstract} 

We find the explicit coordinate transformation which links two exact
cosmological solutions of the brane world which have been recently 
discovered. This means that both solutions are exactly the same with
each other. One of two solutions is described by the motion of a
domain wall in the well-known 5-dimensional Schwarzshild-AdS 
spacetime. Hence, we can easily understand the region covered by
the coordinate used by another solution. 

\end{abstract}



The Randall-Sundrum brane world\cite{RS1,RS2} may give drastic changes
to the conventional gravity theory or cosmology~\footnote{
For earlier works on related scenarios, see
Refs.~\cite{earlier-works}.
}. 
Since the 3-brane world is motivated by the relation between the
$11$-dimensional supergravity and $E_8 \times E_8$ superstring
theory~\cite{Horava}, the drastic changes might be realistic. Hence we
have to seriously think of this scenario. In this respect, the
conventional 4-dimensional theory should be recovered in low energy
limits. This recovery is easily confirmed on a brane with positive 
tension~\cite{Tess,Tama}. Hence for simplicity, hereafter, we will
consider the single brane case with the positive tension. For the
brane world cosmology with negative tension, the issue is still in 
debate~\cite{Tess,Tama,Csaki1,Radion,Radion2,Horowitz}, because 
the fine-tuning and radius stabilization problem seem to be necessary to 
recover the conventional Einstein theory and the single brane is not 
apparently acceptable. 

Recently several authors have found exact cosmological solutions  
in the brane world~\cite{Kraus,Ida,BDEL,Mukohyama,Vollic}
~\footnote{
In this field, a lot of works related have been done so
far~\cite{Csaki1,Kaloper,Nihei,Kim,BDL,Flanagan,Cline,Csaki2,Olive,Misao}.
}.
In one of them, the 3-brane is described as a `domain wall' moving in
5-dimensional black-hole geometries~\cite{Kraus,Ida}. It may be
worth noting that the radiation dominated Friedmann universe is also 
expected by the AdS/CFT correspondence~\cite{Gubser,ADSCFT}. Another
one was given by exactly solving the Einstein equations in the
Gaussian normal coordinate~\cite{BDEL,Mukohyama,Vollic}. 
These global solutions will be important to discuss its stability as
the full spacetime, the gravitational force between two bodies on the
brane and the cosmological perturbation on the brane.

Since each of the two solutions is general enough in each coordinate
system, it is easily expected that both solutions represent the same
spacetime in different coordinate systems. In this brief note, we give
the explicit coordinate transformation from the coordinate used in
Refs.~\cite{Kraus,Ida} to the Gaussian normal coordinate adopted in 
Refs.~\cite{BDEL,Mukohyama,Vollic}. Moreover, we identify the region
which the Gaussian normal coordinate covers.


As stated above, some authors~\cite{Kraus,Ida} considered a
domain-wall moving in the 5-dimensional `Schwarzshild-AdS'(Sch-AdS)
spacetimes~\cite{TBH} with the metric
%
\begin{equation}
 ds^2 = g_{\mu\nu}dx^{\mu}dx^{\nu} = 
	-f(r) dT^2 + \frac{dr^2}{f(r)} + r^2 d\Sigma^2_K,
	\label{eqn:metric-Sch-AdS}
\end{equation}
where $d\Sigma^2_K$ is a metric of a unit three-dimensional sphere,
plane or hyperboloid for $K=+1,0$ or $-1$, respectively, and 
%
\begin{equation}
 f(r) = K + \frac{r^2}{l^2} - \frac{\mu}{r^2}. 
\end{equation}
Here, $l$ ($>0$) and $\mu$ ($\ge 0$ for $K=+1$ or $K=0$, $\ge -l^2/4$ 
for $K=-1$) are constants. The constant $l$ gives the curvature scale
of the bulk spacetime. Hereafter, we denote coordinates on the
three-dimensional manifold (sphere, plane or hyperboloid) by
$x^i$. Therein the domain wall is the 3-brane, whose orbit is given by 
$r=R(T)$, and the bulk spacetime is the 5-dimensional Sch-AdS
spacetime. Assuming the $Z_2$-symmetry which is inspired by the
reduction from M-theory to $E_8 \times E_8$ heterotic string 
theory~\cite{Horava}, Israel's junction condition~\cite{israel} gives
the Friedmann equation on the brane: 
%
\begin{equation}
 \Bigl( \frac{{\dot a}}{a}\Bigr)^2 =
	\frac{8\pi G_N}{3}\rho -\frac{K}{a^2}+
	\frac{\Lambda_4}{3}+\frac{\kappa_5^4}{36}\rho^2+\frac{\mu}{a^4}, 
	\label{eq:fried}
\end{equation}
where the function $R(T)$ representing the orbit is reinterpreted as a
scale factor $a(\tau)$ with the proper time $\tau$, and $\rho$ is
energy density of matter on the domain wall. Here, $\kappa_5$ is the 
$5$-dimensional gravitational constant, $G_N=\kappa_5^4 \lambda
/48\pi$, $\Lambda_4=\kappa_5^4\lambda^2/12-3l^{-2}$ and $\lambda$ is
the vacuum energy on the brane. The deviation from the conventional
Friedmann equation is expressed as the forth and fifth terms in the
right-hand side of Eq. (\ref{eq:fried}). The fifth term comes from the
`electric' part of the 5-dimensional Weyl tensor\cite{Tess}.

The form of Eq. (\ref{eqn:metric-Sch-AdS}) seems simple enough to
handle while a metric derived by several
authors~\cite{BDEL,Mukohyama,Vollic} looks rather complicated. 
However, as already explained above, it is expected that both of
metrics express the same spacetime. 
Yet, the Weyl tensor for the metric derived in
Refs.~\cite{BDEL,Mukohyama,Vollic} becomes zero in the limit of the 
infinite value of the fifth coordinate (or the affine parameter in the 
Gaussian normal coordinate). This implies that the limit of the
infinite affine parameter does not correspond to the Cauchy horizon
since the Weyl tensor cannot be zero except for the conformal infinity
as long as the bulk spacetime is not exactly the anti-deSitter
spacetime. In order to show these explicitly, we transform coordinates
in the metric (\ref{eqn:metric-Sch-AdS}) so that the transformed
coordinate system becomes a Gaussian normal coordinate system based on
a hypersurface given by $r=R(T)$.

For the purpose of the coordinate transformation, let us consider
geodesics which intersect with the hypersurface $r=R(T)$
perpendicularly. Note that these geodesics are spacelike and have zero 
$x^i$-components, provided that the hypersurface $r=R(T)$ is
timelike. First, because of the existence of the Killing field
$(\partial/\partial T)^{\mu}$ in the bulk spacetime, the unit tangent 
vector $u^{\mu}$ of a geodesic should satisfy 
%
\begin{eqnarray}
 g_{\mu\nu}u^{\mu}
	(\partial/\partial T)^{\nu} & = & -E,\nonumber\\
 g_{\mu\nu}u^{\mu}u^{\nu} & = & 1,
\end{eqnarray}
where $E$ is an integration constant. 
Hence, we obtain $u^{\mu}$ as 
%
\begin{equation}
 u^{\mu}\partial_{\mu} = 
	\frac{E}{f(r)}\partial_T \pm\sqrt{f(r)+E^2}\partial_r.
\end{equation}
In the above, we assumed $u^i=0$ 
because we are interested only in geodesics whose tangent have zero
$x^i$-components. 
The trajectory of the geodesic is given by 
%
\begin{equation}
 \frac{dx^{\mu}}{d w} = u^{\mu}, \label{eqn:dxdw=u}
\end{equation}
where $w$ is the affine parameter. For the case of 
$4\mu +l^2(E^2+K)^2>0$, the $r$-component can be integrated as 
%
\begin{equation}
 2 r^2 + l^2(E^2+K) = \sqrt{4l^2\mu+l^4(E^2+K)^2}
	\cosh{[2l^{-1}(\pm w+w_0)]},	\label{eqn:r^2-integral} 
\end{equation}
where $w_0$ is a constant. For the cases of $4\mu +l^2(E^2+K)^2=0$ and
$4\mu +l^2(E^2+K)^2<0$, the $r$-component of Eq.~(\ref{eqn:dxdw=u})
are integrated to give different expressions of $r$ in terms of
$w$. However, the final form of the metric we shall obtain below is
common for all cases. Hence, in the following arguments we show
explicit calculation for the first case only. Let us determine the
constants $E$ and $w_0$ so that the geodesic intersects with the
hypersurface $r=R(T)$ perpendicularly at $T=T_0$ and that the affine 
parameter $w$ is zero on the hypersurface: 
%
\begin{eqnarray}
 u^{\mu} & \propto & g^{\mu\nu}\partial_{\nu}(r-R(T))\qquad
	\mbox{at}\quad T=T_0,r=R(T_0),\nonumber\\
 2 R^2(T_0) + l^2(E^2+K) & = &\sqrt{4l^2\mu+l^4(E^2+K)^2}
	\cosh{(2l^{-1}w_0)},
\end{eqnarray}
These can be solved to give 
%
\begin{eqnarray}
 E & = & E(T_0)  \equiv \pm R'(T_0)
	\sqrt{\frac{f(R(T_0))}{f^2(R(T_0))-{R'}^2(T_0))}},
	\label{eqn:value-E}\\
 w_0 & = & w_0(T_0)\equiv \frac{l}{2}\cosh^{-1}\left[
	\frac{2R^2(T_0)+l^2(E^2(T_0)+K)}{\sqrt{4l^2\mu +l^4(E^2+K)^2}}
	\right],	\label{eqn:value-w0}
\end{eqnarray}
where we have taken a convention such that $\cosh^{-1}X>0$ for $X>1$. 
Note that the set $(T_0,w,x^i)$ can be considered as a coordinate
system.

Next, we can introduce a new coordinate $t$ so that $t$ becomes a
proper time on the hypersurface $r=R(T)$: 
%
\begin{eqnarray}
 \left(\frac{\partial t}{\partial T_0}\right)_{w,x^i} & = &
	\frac{f(a(t))}{\sqrt{f(a(t))+\dot{a}^2(t)}},\\
 \left(\frac{\partial t}{\partial w}\right)_{T_0,x^i} & = & 
 \left(\frac{\partial t}{\partial x^i}\right)_{T_0,w} = 0,
\end{eqnarray}
where $a(t)=R(T_0(t))$. Note that the function $a(t)$ is well-defined
in a whole coordinate patch as well as on the hypersurface $r=r(T)$,
or $w=0$, since both $t$ and $T_0$ are constant along the geodesic. 
In this coordinate, $E(T_0)$ given by Eq.~(\ref{eqn:value-E}) has a
simple expression 
%
\begin{equation}
 E(T_0) = \pm\dot{a}(t),\label{eqn:value-E2}
\end{equation}
where a dot denotes $(\partial/\partial t)_{w,x^i}$. Hence, by
substituting Eqs.~(\ref{eqn:value-E2}) and (\ref{eqn:value-w0}) into
Eq.~(\ref{eqn:r^2-integral}), we can rewrite the original coordinate
$r$ in terms of new coordinates $t$ and $w$ as 
%
\begin{eqnarray}
 r^2 = \varphi(t,w)a^2(t), \label{eqn:r^2}
\end{eqnarray}
where
%
\begin{equation}
 \varphi(t,w) = \cosh{(2l^{-1}w)} + \frac{l^2}{2}(H^2+Ka^{-2})
	(\cosh{(2l^{-1}w)} - 1) 
	\pm \sqrt{1 + l^2(H^2+Ka^{-2}-\mu a^{-4})}\sinh{(2l^{-1}w)}. 
	\label{eqn:varphi}
\end{equation}
Here $H(t)$ is defined by 
%
\begin{equation}
 H(t) = \frac{\dot{a}(t)}{a(t)}.
\end{equation}

Now let us confirm that the new coordinate system $(t,w,x^i)$ is
actually a Gaussian normal coordinate system. For this purpose, it is
sufficient to show that 
%
\begin{equation}
 dw = g_{\mu\nu}u^{\mu}dx^{\nu}, 
\end{equation}
or
%
\begin{eqnarray}
 \left(\frac{\partial w}{\partial T}\right)_{r,x^i} & = & 
	\mp\dot{a}(t),\nonumber\\
 \left(\frac{\partial w}{\partial r}\right)_{T,x^i} & = & 
	\pm\frac{\sqrt{f(r)+\dot{a}^2(t)}}{f(r)}.
	\label{eqn:dw},\nonumber\\
 \left(\frac{\partial w}{\partial x^i}\right)_{T,r} & = & 0
\end{eqnarray}
is integrable. 
The integrability condition $ddw=0$ is equivalent to 
%
\begin{equation}
 \left(\frac{\partial t}{\partial T}\right)_{r,x^i}/
	\left(\frac{\partial t}{\partial r}\right)_{T,x^i}
 = -\frac{f(r)\sqrt{f(r)+\dot{a}^2}}{\dot{a}}.
\end{equation}
This condition is easily confirmed by using the relation
%
\begin{equation}
 \left(\frac{\partial t}{\partial T}\right)_{r,x^i}/
	\left(\frac{\partial t}{\partial r}\right)_{T,x^i}
 = -\left(\frac{\partial r}{\partial w}\right)_{t,x^i}/
	\left(\frac{\partial T}{\partial w}\right)_{t,x^i},
\end{equation}
and 
%
\begin{eqnarray}
 \left(\frac{\partial T}{\partial w}\right)_{t,x^i}
 & = & \left(\frac{\partial T}{\partial w}\right)_{T_0,x^i}
 = u^T = \pm\frac{\dot{a}(t)}{f(r)},\nonumber\\
 \left(\frac{\partial r}{\partial w}\right)_{t,x^i}
 & = & \left(\frac{\partial r}{\partial w}\right)_{T_0,x^i}
 = u^r = \pm \sqrt{f(r)+\dot{a}^2}. \label{eqn:w-derivative}
\end{eqnarray}

Finally, we can calculate $(\partial r/\partial t)_{w,x^i}$ from
(\ref{eqn:r^2}) as well as $(\partial T/\partial w)_{t,x^i}$ and
$(\partial r/\partial w)_{T,x^i}$ from (\ref{eqn:w-derivative}). 
Moreover, from Eq.~(\ref{eqn:dw}), 
%
\begin{equation}
 \left(\frac{\partial T}{\partial t}\right)_{w,x^i}/
	\left(\frac{\partial r}{\partial t}\right)_{w,x^i}
 =  - \left(\frac{\partial w}{\partial r}\right)_{T,x^i}/
	\left(\frac{\partial w}{\partial T}\right)_{r,x^i}
 = \mp\frac{\sqrt{f(r)+\dot{a}^2}}{f(r)\dot{a}(t)}.
\end{equation}
Thus, it is easily shown that 
%
\begin{equation}
 ds^2 = -\frac{\psi^2(t,w)}{\varphi(t,w)}dt^2 + 
	\varphi(t,w)a(t)^2d\Sigma^2_K + dw^2,\label{eqn:BDEL-metric}
\end{equation}
where
%
\begin{eqnarray}
 \psi(t,w) & \equiv & \varphi + \frac{1}{2H}
	\left(\frac{\partial\varphi}{\partial t}\right)_{w,x^i}
	\nonumber\\
 & = & \cosh{(2l^{-1}w)} + \frac{l^2}{2}(H^2+\dot{H})
	(\cosh{(2l^{-1}w)}-1) \pm
	\frac{1+\frac{l^2}{2}(2H^2+\dot{H}+Ka^{-2})}
	{\sqrt{1+l^2(H^2+Ka^{-2}-\mu a^{-4})}}\sinh{(2l^{-1}w)}.
\end{eqnarray}
By defining $k$ and $C$ by
%
\begin{eqnarray}
 k & = & 2l^{-1},\nonumber\\
 C & = & -l^2\mu,
\end{eqnarray}
the functions $\psi(t,w)$ and $\varphi(t,w)$ are rewritten as
%
\begin{eqnarray}
 \psi(t,w) & = & \cosh{(kw)} + 2k^{-2}(H^2+\dot{H})
	(\cosh{(kw)}-1) \pm
	\frac{1+2k^{-2}(2H^2+\dot{H}+Ka^{-2})}
	{\sqrt{1+4k^{-2}(H^2+Ka^{-2})+Ca^{-4})}}\sinh{(kw)},\nonumber\\
 \varphi(t,w) & = & \cosh{(kw)} + 2k^{-2}(H^2+Ka^{-2})
	(\cosh{(kw)} - 1) 
	\pm \sqrt{1+4k^{-2}(H^2+Ka^{-2})+Ca^{-4})}\sinh{(kw)}.
	\label{eqn:psi-varphi}
\end{eqnarray}
Using the Friedmann equation of Eq. (\ref{eq:fried}) and setting 
$C={\cal C}\kappa_5^4 l^2$, we can see that the metric 
(\ref{eqn:BDEL-metric}) with (\ref{eqn:psi-varphi}) is equivalent to 
the expression obtained in Ref. \cite{BDEL}. 
As shown in Refs.~\cite{BDEL,Mukohyama,Vollic} the lower signs should
be taken in all equations if we glue two copies of the region $w\ge 0$
to obtain the brane-world with {\it positive} tension. Thus,
hereafter, we take the lower signs.


{}From now on we show where the Gaussian normal coordinate covers in the
Sch-AdS spacetime. For simplicity, we will consider only $\Lambda_4=0$
cases. 
We concentrate on the $\mu\neq 0$ cases since the
$\mu=0$ case is easier and can be understood in a similar way. Note
that the metric (\ref{eqn:metric-Sch-AdS}) has an event horizon at
$r=r_h$, where $r_h$ is given by 
%
\begin{equation}
 r_h^2 = \frac{l^2}{2}(\sqrt{K^2+4l^{-2}\mu}-K).
\end{equation}
In the following arguments we shall show that the Gaussian normal
coordinate system $(t,w,x^i)$ covers the region beyond the event horizon and
the wormhole.

First, let us consider the case when the condition
%
\begin{equation}
 (H^2a^2+K)^2 + 4l^{-2}\mu > 0
\end{equation}
is satisfied. For $K=+1$, this condition is automatically
satisfied since $\mu\ge 0$ for the bulk {\it black hole} 
spacetimes. This conditions is automatically satisfied
also when $K=0$ and $Ha\ne 0$. In this case, it is easily shown that 
%
\begin{equation}
 r^2 = \varphi(t,w)a(t)^2 \to \infty \qquad (w\to\infty)
\end{equation}
and that 
%
\begin{equation}
 r^2 = \varphi(t,w)a(t)^2 \ge r_{min}^2,\qquad (w\ge 0)
	\label{eqn:r-lowerlimit}
\end{equation}
where
%
\begin{equation}
 r_{min}^2 = \frac{l^2}{2}
	\left[\sqrt{(H^2a^2+K)^2+4l^{-2}\mu}-(H^2a^2+K)\right].
\end{equation}
Note that $r_{min}(t)$ has been determined by the variation of $w$
under a fixed $t$. The equality in Eq.(\ref{eqn:r-lowerlimit}) holds
at $w=w_{min}(t)$, where $w_{min}(t)$ ($>0$) is given by 
%
\begin{equation}
 \cosh{(2l^{-1}w_{min}(t))} =
 \frac{2l^{-2}a^2+H^2a^2+K}{\sqrt{(H^2a^2+K)^2+4l^{-2}\mu}}.
\end{equation}
We can easily show that $r_{min}\to 0$ and $w_{min}\to 0$ as $a\to 0$
by using the Friedmann equation on the brane. 
On the other hand, there is another minimum $r_{min}^*(w)$ which 
is determined by the variation of $t$ under a fixed $w$. The orbit 
of $r_{min}^*(w)$ is the same as $\psi (t,w)=0$ because $\psi$ 
is proportional to $\partial_t r^2$. These minimum will be important 
when we draw the conformal diagram. 

Now it is easily shown by using the lower-limit of $\mu$ ($\mu\ge 0$
for $K=+1$ and $K=0$, $\mu\ge -l^2/4$ for $K=-1$, and we are
considering the $\mu\ne 0$ case) that $r_{min}^2\le r_h^2$. The
equality holds if and only if $Ha=0$. Hence, the coordinate
$(t,w,x^i)$ actually covers the region beyond the event horizon and
the wormhole. 

Next, let us consider the case when $K=-1$ and the condition
%
\begin{equation}
 (H^2a^2-1)^2 + 4l^{-2}\mu = 0
\end{equation}
is satisfied. In this case, $r^2$ approaches to a finite value in the
limit $w\to\infty$: 
%
\begin{equation}
 r^2 = \varphi(t,w)a(t)^2 \to \frac{l^2}{2}(1-H^2a^2) < \frac{l^2}{2} 
	\qquad (w\to\infty).
\end{equation}
On the other hand, 
%
\begin{equation}
 r_h^2 > \frac{l^2}{2}.
\end{equation}
Therefore, the coordinate $(t,w,x^i)$ reaches the region beyond the
event horizon.

Finally, let us consider the case of $K=-1$ and the time when condition 
%
\begin{equation}
 (H^2a^2-1)^2 + 4l^{-2}\mu < 0
\end{equation}
is satisfied. In this case, there exists a value of $w$ ($>0$) such
that $r^2=0$. Note that $r=0$ corresponds to a singularity inside the
horizon since tetrad components of the Weyl tensor for the metric 
(\ref{eqn:metric-Sch-AdS}) diverge at $r=0$. Thus, the coordinate
$(t,w,x^i)$ covers the region inside the event horizon and reaches the
singularity inside the horizon.

\begin{figure}
 \begin{center}
  \epsfile{file=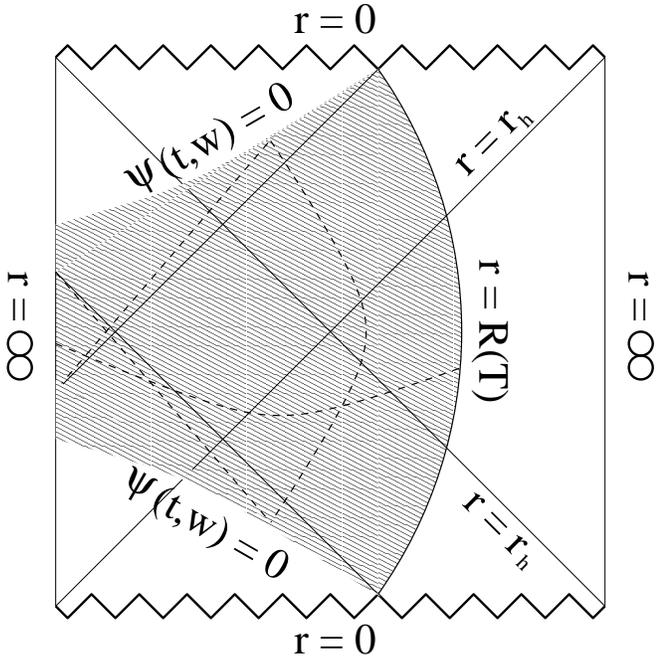,scale=0.5}
 \end{center}
\caption{
The global structure for $\mu>0$, $K=+1$. The world volume of the
brane starts at $r=0$ and ends at $r=0$. The function $r_{min}(t)$
increases from zero to $r_h$, and decreases to zero. Accordingly, the
hypersurface $w=w_{min}(t)$ starts at $r=0$ with the brane, passes
through the bifurcating point of the horizon, and ends at $r=0$ with
the brane. Thus, the region covered by the Gaussian normal coordinate
should be the shaded region. Two dashed lines in the figure are a
constant-$t$ hypersurface and a constant-$w$ hypersurface. 
}
\label{fig:Fig1}
\end{figure}

\begin{figure}
 \begin{center}
  \epsfile{file=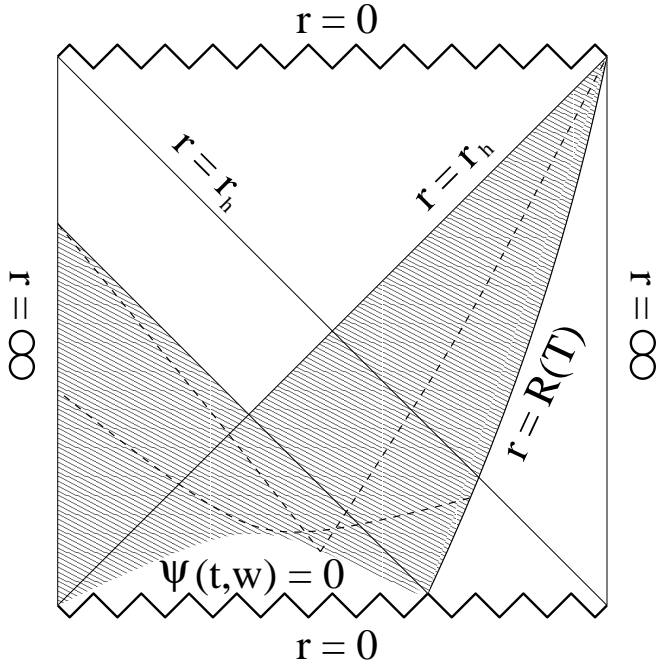,scale=0.5}
 \end{center}
\caption{
The global structure for $\mu>0$, $K=0,-1$. The world volume of the
brane starts at $r=0$ and ends at $r=\infty$. The function
$r_{min}(t)$ starts with the value zero and ends with the value
$r_{min}(\infty)$, where $r_{min}(\infty)=r_h$ for $K=0$ and
$r_{min}(\infty)=(l^2\mu)^{1/4}<r_h$ for $K=-1$. (We have assumed that
$\Lambda_4=0$ and $\rho a^2\to 0$ in the limit $a\to\infty$.) Thus,
the region covered by the Gaussian normal coordinate should be the
shaded region. Note that hypersurfaces $w=w_{min}(t)$ and
$\psi(t,w)=0$ start at the point where the brane starts, and end at
the point where $r=r_{min}(\infty)$ comes in contact with the
horizon. Two dashed lines in the figure are a constant-$t$
hypersurface and a constant-$w$ hypersurface.
}
\label{fig:Fig2}
\end{figure}

\begin{figure}
 \begin{center}
  \epsfile{file=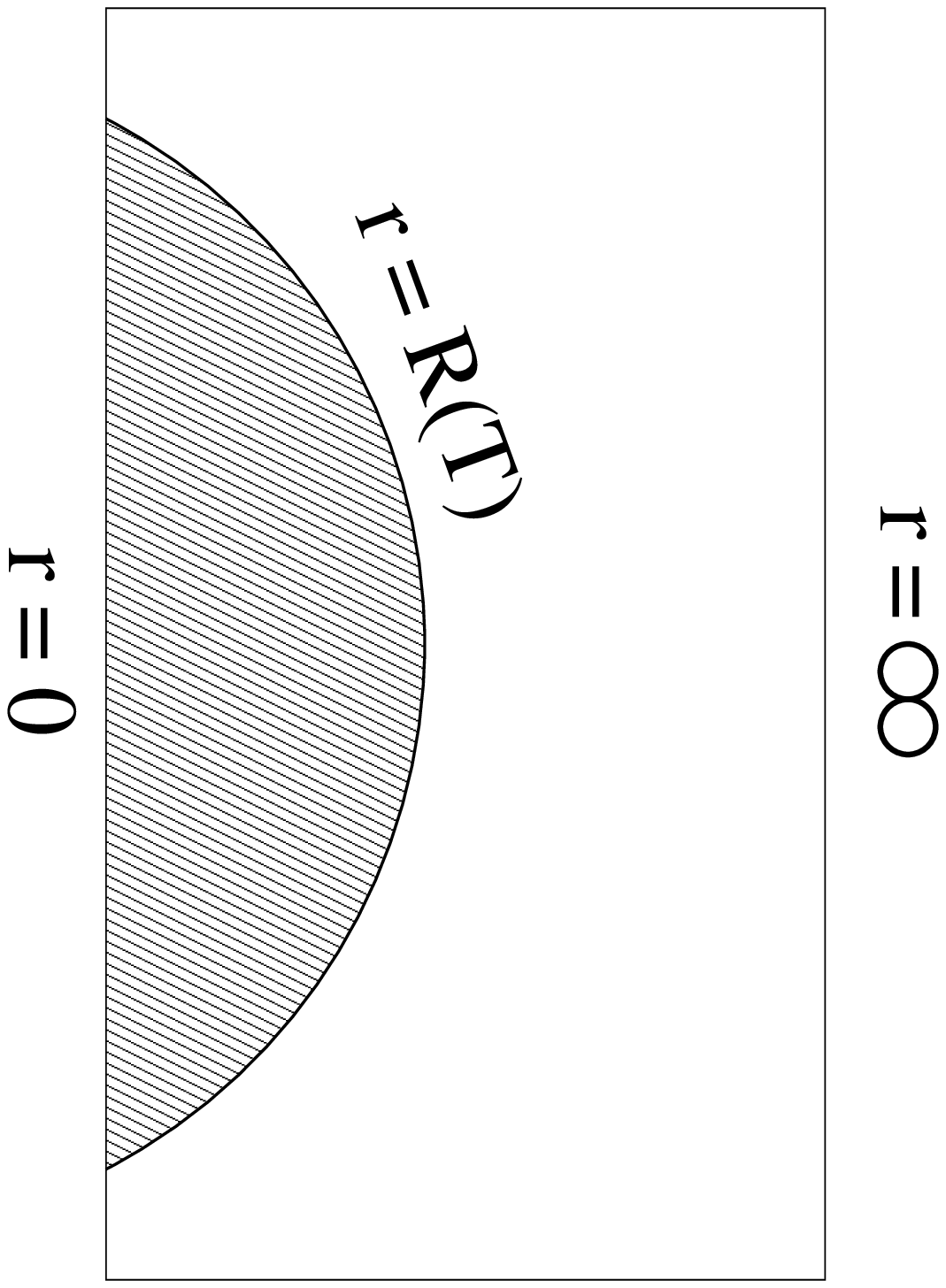,scale=0.5}
 \end{center}
\caption{
The global structure for $\mu=0$, $K=+1$. The world volume of the
brane starts at $r=0$ and ends at $r=0$. The minimum of $r$ for a
fixed $t$ is always zero. Thus, the region covered by the Gaussian
normal coordinate should be the shaded region. 
}
\label{fig:Fig3}
\end{figure}

\begin{figure}
 \begin{center}
  \epsfile{file=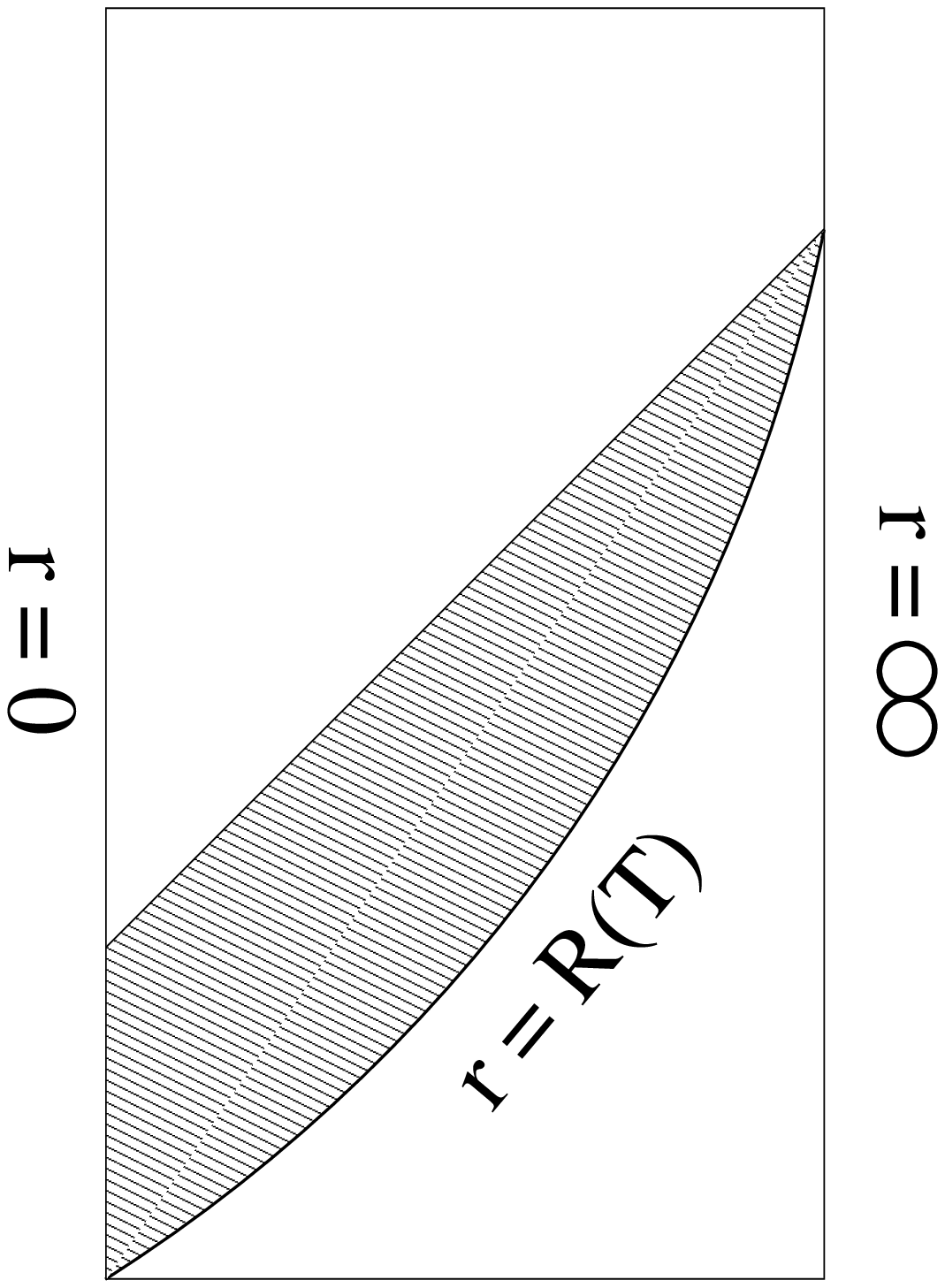,scale=0.5}
 \end{center}
\caption{
The global structure for $\mu=0$, $K=0,-1$. The world volume of the
brane starts at $r=0$ and ends at $r=\infty$. The minimum of $r$ for a
fixed value of $t$ is always zero. Thus, the region covered by the
Gaussian normal coordinate should be the shaded region. 
}
\label{fig:Fig4}
\end{figure}

\begin{figure}
 \begin{center}
  \epsfile{file=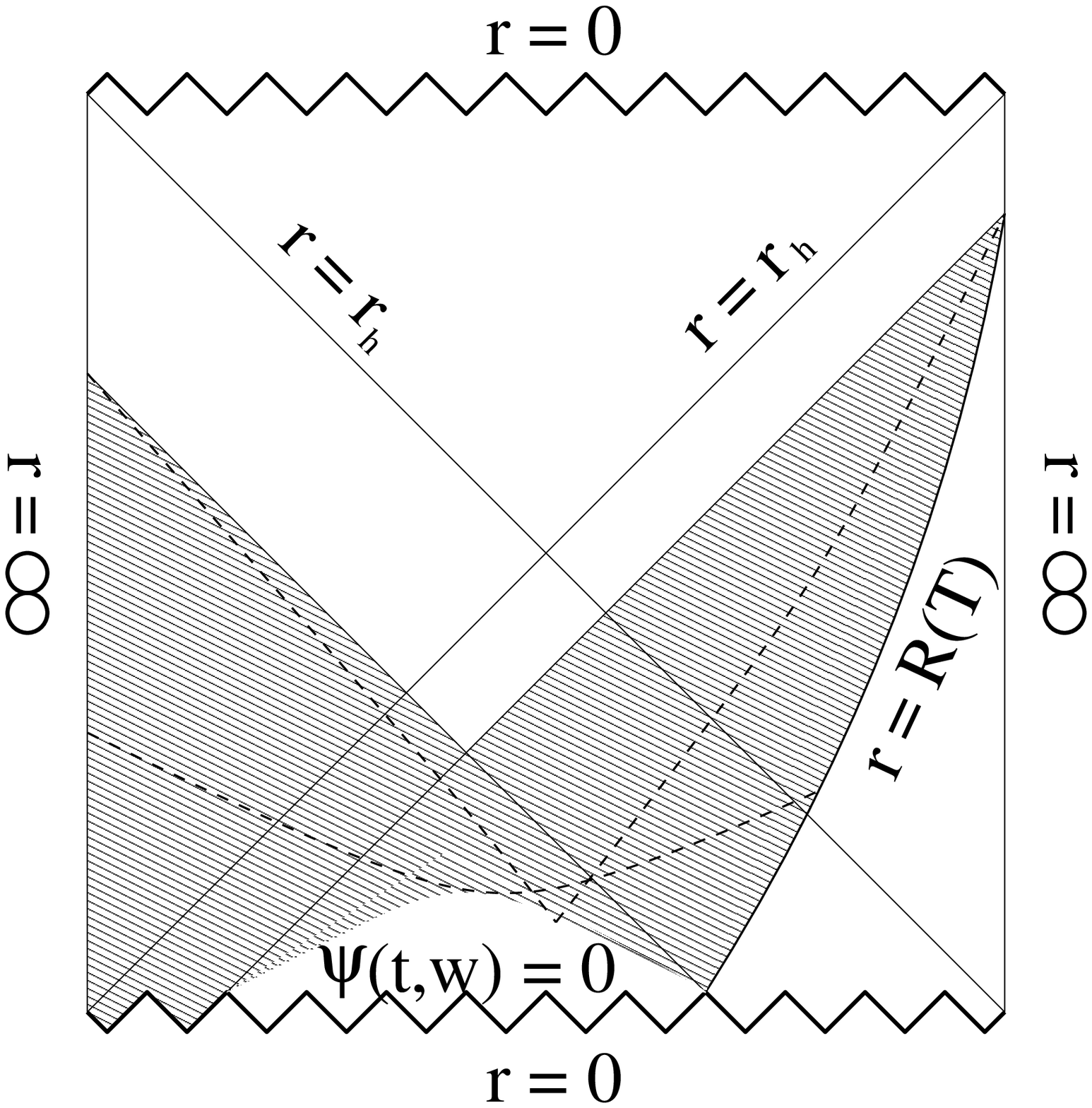,scale=0.5}
 \end{center}
\caption{
The global structure for $\mu<0$, $K-1$. The world volume of the brane
starts at $r=0$ and ends at $r=\infty$. The minimum of $r$ for a fixed
$t$ becomes zero in the limits $a\to 0$ and $a\to\infty$. Thus, the
region covered by the Gaussian normal coordinate should be the shaded 
region. Note that hypersurfaces $w=w_{min}(t)$ and $\psi(t,w)=0$ start
at the point where the brane starts, and end at $r=0$. Two dashed
lines in the figure are a constant-$t$ hypersurface and a constant-$w$ 
hypersurface. 
}
\label{fig:Fig5}
\end{figure}

{}Now it is easy to see the global structure. Figure \ref{fig:Fig1} is
for $\mu>0$, $K=+1$. Figure \ref{fig:Fig2} is for $\mu>0$,
$K=0,-1$. Figures \ref{fig:Fig3} and \ref{fig:Fig4} are for $\mu=0$,
$K=+1$ and for $\mu=0$, $K=0,-1$, respectively. Figure \ref{fig:Fig5}
is for $\mu<0$, $K=-1$. In these figures, the region covered by the
Gaussian normal coordinate is expressed as the shaded region. Note
that in Figures \ref{fig:Fig1}, \ref{fig:Fig2} and \ref{fig:Fig5}, 
for a non-zero small value of
$w$, $\partial_t$ should be past directed for small $t$, turns to 
the opposite direction at a time $t_*$ when the orbit of 
$\partial_t$ reaches the $r_{min}^*(w)$, and future directed for large $t$.

{}To obtain these figures, we have used the following two facts. First,
the hypersurface $t=t_0$ is always spacelike, and should come in
contact with the hypersurface $r=r_{min}(t_0)$ at $w=w_{min}(t_0)$, 
where $t_0$ is a constant. Secondly, for a fixed $w$, $r\to\infty$ in
the limits $a\to 0$ and $a\to\infty$, providing some reasonable
assumptions (eg. $\Lambda_4=0$ and $\rho a^2\to 0$ in the limit
$a\to\infty$). This means that the constant-$t$ hypersurface should
become null in these limits.


In this brief note, we have given the coordinate transformation between 
the metric of Eq.~(\ref{eqn:metric-Sch-AdS}) and of
Eq.~(\ref{eqn:BDEL-metric}).  As a result, we could see the region
where the Gaussian normal coordinate covers. For general cases of
single brane with the positive tension and $\mu \neq 0$, the
coordinate $w$ labelling the extra dimension does not terminate at the
`Cauchy horizon' and goes beyond the event horizon. For some cases,
the coordinate goes through the wormhole and reaches another domain of 
communication where we can define the total energy well~\cite{AD}. The
energy should be $\mu$ for cases which we considered and we can prove
the positivity on the slice without naked singularities at least $K=0$
cases~\cite{Bogo,Wool}. 
This might argue us to consider general issues such as the positivity
of the ADM energy~\cite{PET}, cosmic no-hair~\cite{Wald}, singularity
theorem~\cite{GR} and so on. According to Ref.~\cite{Tess}, the
effective 4-dimensional Einstein equation is given by 
%
\begin{equation}
 {}^{(4)}G_{\mu\nu}= 
	-\Lambda_4 q_{\mu\nu}+8\pi G_N T_{\mu\nu}
	+\kappa_5^4 \Pi_{\mu\nu}-E_{\mu\nu},
	\label{eq:effective}
\end{equation}
%
where $T_{\mu\nu}$ is the energy-momentum tensor on the brane,
$\Pi_{\mu\nu}$ is a quadratic term of $T_{\mu\nu}$,
and $E_{\mu\nu}={}^{(5)}C_{\mu\alpha\nu\beta}n^\alpha n^\beta$ is the 
`electric' part of the 5-dimensional Weyl tensor. It is
important to clarify whether the right-hand side of
Eq. (\ref{eq:effective}) satisfies the local energy
condition~\cite{GR} or not, because it is closely
related to many interesting subjects such as stability
of the Minkowski spacetime. For the Sch-AdS spacetime,
$E_{00}\sim -\mu/a^4$ and then $-E_{00} \geq 0$ is guaranteed 
by the above argument. 
This indicates the tendency such that the right-hand side satisfies 
the local energy condition. However, we know that 
the perturbation analysis on the Randall \& Sundrum brane 
world implies $E_{00} \geq 0$~\cite{Tess2}. Hence, we cannot 
quickly give the definite answer on the signature. This is crucial 
problem for general issues.

\begin{acknowledgments}

SM would like to thank Professor W. Israel for continuing
encouragement and Professor L. Kofman for his warm hospitality in
Canadian Institute for Theoretical Astrophysics. TS would like to 
thank D. Langlois, M. Sasaki and T. Tanaka for discussions. SM's work is
supported by the CITA National Fellowship and the NSERC operating
research grant. TS's work is supported by JSPS Postdoctal Fellowship for 
Research Abroad. KM's work  is supported in part by Monbusho
Grant-in Aid for Specially Promoted Research No.~08102010.

\end{acknowledgments}


\begin{references}
\bibitem{RS1}
L.~Randall and R.~Sundrum, Phys. Rev. Lett. {\bf 83},  3370 (1999)
[hep-ph/9905221]. 
\bibitem{RS2}
L.~Randall and R.~Sundrum, Phys.Rev.Lett. {\bf 83}, 4690 (1999)
[hep-th/9906064].
\bibitem{earlier-works}
V.~A.~Rubakov and M.~E.~Shaposhnikov, Phys. Lett. {\bf 152B}, 136
(1983);\\
M.~Visser, Phys. Lett. {\bf B159}, 22 (1985);\\
K.~Akama, Prog. Theor. Phys. {\bf 78}, 184 (1987). 
\bibitem{Horava}
P.~Horava and E.~Witten, Nucl. Phys. {\bf 460}, 506(1996).
\bibitem{israel}
W. Isarel, Nuovo Cim. {\bf 44B}, 1 (1966). 
\bibitem{Tess}
T.~Shiromizu, K.~Maeda and M.~Sasaki, gr-qc/9910076.
\bibitem{Tama}
J.~Garriga and T.~Tanaka, hep-th/9911055.
\bibitem{Csaki1}
C.~Csaki, M.~Graesser, C.~Kold and J.~Terning, hep-ph/9906513.
\bibitem{Radion}
C.~Csaki, M.~Graesser, L.~Randal and J.~Terning, hep-ph/9911406.
\bibitem{Radion2}
P.~Kanti, I.~I.~Kogan, K.~A.~Olive and M.~Pospelov, 
hep-ph/9912266.
\bibitem{Horowitz}
R.~Emparan, G.~T.~Horowitz and R.~C.~Myers, hep-th/9911043.
\bibitem{Kraus}
P. Kraus, JHEP {\bf 9912}, 011 (1999) [hep-th/9910149]. 
\bibitem{Ida}
D.~Ida, gr-qc/9912002.
\bibitem{BDEL}
P.~Bin\'{e}truy, C.~Deffayet, U.~Ellwanger and D.~Langlois,
hep-th/9910219. 
\bibitem{Mukohyama}
S.~Mukohyama, hep-th/9911165, to appear in Phys. Lett. {\bf B}.
\bibitem{Vollic}
D.~N.~Vollick, hep-th/9911181.
\bibitem{Kaloper}
N.~Kaloper, hep-th/9905210.
\bibitem{Nihei}
T.~Nihei, hep-ph/9905487.
\bibitem{Kim}
H.~B.~Kim  and H.~D.~Kim, hep-th/9909053. 
\bibitem{BDL}
P.~Binetruy, C.~Deffayet and D.~Langlois, hep-th/9905012.
\bibitem{Flanagan}
E.~E.~Flanagan, S.~H.~H.~Tye, I.~Wasserman, hep-ph/9910498. 
\bibitem{Cline}
J.~M.~Cline, C.~Grojean and G.~Servant, hep-ph/9906523.
\bibitem{Csaki2}
C.~Csaki, M.~Graesser, L.~Randall and J.~Terning, hep-ph/9911406. 
\bibitem{Olive}
P.~Kanti, I.~I.~Kogan, K.~A.~Olive and M.~Prospelov, hep-ph/9909481.
\bibitem{Misao}
J.~Garriga and M.~Sasaki, hep-th/9912118. 
\bibitem{Gubser}
S.~S.~Gubser, hep-th/9912001.
\bibitem{ADSCFT}
For AdS/CFT correspondece, \\
J.~Maldacena, Adv. Theor. Math. Phys., {\bf 2}, 231(1998);\\
O.~Aharony, S.~S.~Gubser, J.~Maldacena, H.~Ooguri 
and Y.~Oz, hep-th/9905111 and reference therein 
\bibitem{TBH}
D.~Birmingham, Class. Quant. Grav. {\bf 16}, 1197(1999).
\bibitem{AD}
L.~F.~Abbot and S.~Deser, Nucl. Phys. {\bf 195}, 76(1982);\\
A.~Ashtekar and A.~Magnon, Class. Quantum Grav. {\bf 1}, L39(1984);\\
A.~Ashtekar and S.~Das, hep-th/9911230.
\bibitem{Bogo}
G.~W.~Gibbons, S.~W.~Hawking, G.~T.~Horowitz and M.~J.~Perry, Commun. Math. 
Phys. {\bf 88}, 295(1983).
\bibitem{Wool}
E.~Woolgar, Class. Quantum Grav. {\bf 11},1881(1994). 
\bibitem{PET}
R. Schoen and S. T. Yau, Commun. Math. Phys. {\bf 79}, 23(1981) \\
E. Witten, Commun. Math. Phys. {\bf 80}, 381(1981).
\bibitem{Wald}
R.~M.~Wald, Phys. Rev. {\bf D28}, 2118(1983).
\bibitem{GR}
For example, R. M. Wald, {\it General Relativity},
(Univ. Chicago Press, Chicago, 1984).
\bibitem{Tess2}
M.~Sasaki, T.~Shiromizu and K.~Maeda, hep-th/9912233.
\end{references}
\end{document}